\documentclass[letterpaper, 10 pt, conference]{ieeeconf}  % Comment this line out if you need a4paper

\IEEEoverridecommandlockouts                              % This command is only needed if 
\usepackage{cite}% cite the reference
\usepackage{indentfirst}
\usepackage{graphicx}
\usepackage{amssymb} %beautiful large equal \geqslant, small equal \leqslant ; not beautiful large equal  \geq, small equal \leq
\usepackage{amsfonts}
\usepackage{amsmath}
\usepackage{cases}
\usepackage{algpseudocode}
\usepackage{subfigure}
\usepackage{mathrsfs}
\usepackage{makecell}
\usepackage[linesnumbered,boxed,ruled,commentsnumbered]{algorithm2e}
\SetKwRepeat{Do}{do}{while}%
\usepackage[ruled]{algorithm2e}
\usepackage{multirow}
\usepackage{url}
\usepackage{amssymb}
\usepackage{color}
\usepackage{textcomp}
\usepackage{afterpage}

\usepackage{url}

\title{\LARGE \bf
Distributed Charging Coordination of Electric Trucks with\\ Limited Charging Resources
}

\author{Ting Bai$^{1}$, Yuchao Li$^{2}$, Karl H. Johansson$^{1}$,~\IEEEmembership{Fellow, IEEE}, and Jonas M{\aa}rtensson$^{1}$% <-this % stops a space
\thanks{This work is supported in part by the Swedish Research Council Distinguished Professor (Grant Number: 2017-01078), the Knut and Alice Wallenberg Foundation, and the Swedish Strategic Research Foundation CLAS (Grant Number: RIT17-0046).}% <-this % stops a space
\thanks{$^{1}$T. Bai, K. H. Johansson, and J. M{\aa}rtensson are with the Integrated Transport Research Laboratory, the Division of Decision and Control Systems, and the Digital Futures, KTH Royal Institute of Technology, 100 44 Stockholm, Sweden. E-mails: \{{\tt\small tingbai, kallej, jonas1\}@kth.se}}
\thanks{$^{2}$Y. Li is with the School of Computing and Augmented Intelligence, Arizona State University, AZ-85281 Tempe, the United States. E-mail: {\tt\small yuchaoli@asu.edu}}
}

\begin{document}

\maketitle

%%%%%%%%%%%%%%%%%%%%%%%%%%%%%%%%%%%%%%%%%%%%%%%%%%%%%%%%%%%%%%%%%%%%%%%%%%%%%%%%
\begin{abstract}
Electric trucks usually need to charge their batteries during long-range delivery missions, and the charging times are often nontrivial. As charging resources are limited, waiting times for some trucks can be prolonged at certain stations. To facilitate the efficient operation of electric trucks, we propose a distributed charging coordination framework. Within the scheme, the charging stations provide waiting estimates to incoming trucks upon request and assign charging ports according to the first-come, first-served rule. Based on the updated information, the individual trucks compute where and how long to charge whenever approaching a charging station in order to complete their delivery missions timely and cost-effectively. We perform empirical studies for trucks traveling over the Swedish road network and compare our scheme with the one where charging plans are computed offline, assuming unlimited charging facilities. It is shown that the proposed scheme outperforms the offline approach at the expense of little communication overhead.
\end{abstract}

%%%%%%%%%%%%%%%%%%%%%%%%%%%%%%%%%%%%%%%%%%%%%%%%%%%%%%%%%%%%%%%%%%%%%%%%%%%%%%%%
\section{Introduction}\label{Section I}
Vehicle electrification has emerged as a global trend to cope with climate change and energy shortages~\cite{bao2021global}. As road freight transportation accounts for sizable portions of emissions both in Europe~\cite{siskos2019assessing,10209062} and around the globe~\cite{ritchie2023sector}, replacing diesel-engine trucks with electric ones can bring major benefits. In this endeavor, various tax reductions and subsidiary policies have been introduced \cite{yan2018economic,bai2023third} to increase the market shares of electric trucks. Still, however, several challenges remain to hinder their wide adoption. 

One such problem is known as \emph{range anxiety}~\cite{yong2023electric}, which refers to the unease of drivers due to insufficient battery energy for reaching the intended destinations. This is particularly true for trucks with long-range delivery missions, where even fully charged batteries can not cover the whole range. Therefore, it is often necessary to plan where and how long to recharge the truck within a given collection of charging stations. To date, there have been extensive works developing viable charging strategies. For instance, \cite{zahringer2022time} and \cite{10147895} investigated the optimal planning of charging stops for electric trucks, which also integrate with the mandatory rest schedules of drivers. Some other works, such as~\cite{huber2015long,erdelic2019survey,kobayashi2011route}, transformed the optimal charging problem into the well-known vehicle routing problem, where optimality can refer to consuming the least amount of energy, cost, or time. 

In all the works mentioned earlier, charging ports are assumed to be available whenever a truck arrives at a charging station. However, in reality, this could be hard to guarantee, given the limited charging resources at every station, the long charging procedure, as well as the unknown charging plans of others. As a result, drivers may find themselves caught in prolonged waiting queues, leading to larger labor costs, and potential violation of the delivery deadlines. To address the tension between the limited charging infrastructure and the desired growing number of electric trucks, research emphasizing the coordination of charging strategies has also gained increasing attention in recent years. The coordination can be achieved by trucks, stations, or both parties combined, leading to various architectures.

A considerable amount of literature researches how trucks compute their charging strategies with little effort required on the station side. These works can be classified according to the objects they optimize, including minimizing the additional waiting time at stations~\cite{yang2013charge,qin2011charging}, the total travel time of the journey~\cite{del2016smart}, and the charging and travel times combined with the charging cost~\cite{moghaddam2017smart}. In these works, stochastic arrival processes of vehicles are assumed or synthetic trip data are leveraged in modeling the queuing system at charging stations. Some research treats the charging coordination problem from the perspective of stations that plan the charging schedules for trucks. The goal is to optimize the charging facility usage~\cite{gusrialdi2014scheduling}, distribute charging demands among stations~\cite{gusrialdi2017distributed}, or avoid peak load in the smart grid~\cite{nguyen2014charging}, to name only a few. A few studies have also investigated efficient approaches to coordinating charging behaviors holistically. For example,~\cite{tang2020congestion} proposed a heuristic algorithm to seek optimal charging scheduling by maximizing the social welfare of electric vehicles, charging stations, and power plants. Similar centrally coordinated methods typically require extensive information communication and full control of each component involved in the network.  

In this work, we present a two-layer charging coordination framework to facilitate the charging planning of trucks and to mitigate congestion in charging stations. Within our scheme, trucks compute their individual charging plans without informing each other. Instead, it is based upon information exchanged with nearby stations. The stations schedule charging orders according to the first-come, first-served rule while providing local information to trucks upon request for collective good. Although our scheme shares in spirit some features of the methods in the literature, it highlights the following properties: 
\begin{itemize}
    \item[i)] It is fully distributed and applies to cases where all parties do not belong to the same fleet;
    \item[ii)] It is light-weight, involving only simple information exchange between trucks and stations;
    \item[iii)] It is adaptive, enabling changes in the charging plans of trucks when needed via real-time computation.  
\end{itemize}

The remainder of this paper is structured as follows. Section~\ref{Section II} provides an overview of the two-layer coordination framework consisting of trucks and charging stations. The scheduling mechanism and waiting time computation performed by charging stations are introduced in Section~\ref{Section III}. Then, in Section~\ref{Section IV}, we present the charging planning computation of individual trucks when they receive the information provided by nearby stations. The results of numerical studies based on realistic data are presented in Section~\ref{Section V}, followed by concluding remarks and future research directions in Section~\ref{Section VI}.

\section{Overview of the Coordination Framework}\label{Section II}
In this section, we provide an overview of the proposed coordination framework. It is applied to a collection of charging stations with communication capabilities to trucks and a large number of trucks aiming to minimize their own transport costs. We emphasize the information exchanged between the stations and trucks, the computation executed by them, as well as the instances at which the communication and computation are carried out. Further details will be given in the subsequent sections.

Throughout we assume the following conditions hold: 
\begin{itemize}
    \item[a)] There is a fixed number of charging stations, each with a fixed amount of charging ports. 
    \item[b)] For all the trucks traversing the road network, their routes are pre-planned and fixed, and heading to charging stations may lead to detours.
    \item[c)] All the charging stations and trucks are operating for their own interests.
\end{itemize}
   
In particular, conditions a) and b) simplify the computations required for both charging stations and trucks. Our proposed framework is tailored for problems where condition c) holds, but may also be used in cases where either part of or all the stations and/or trucks operate in collaboration.

\begin{figure}[t!]
     \centering
     \includegraphics[width=0.88\linewidth]{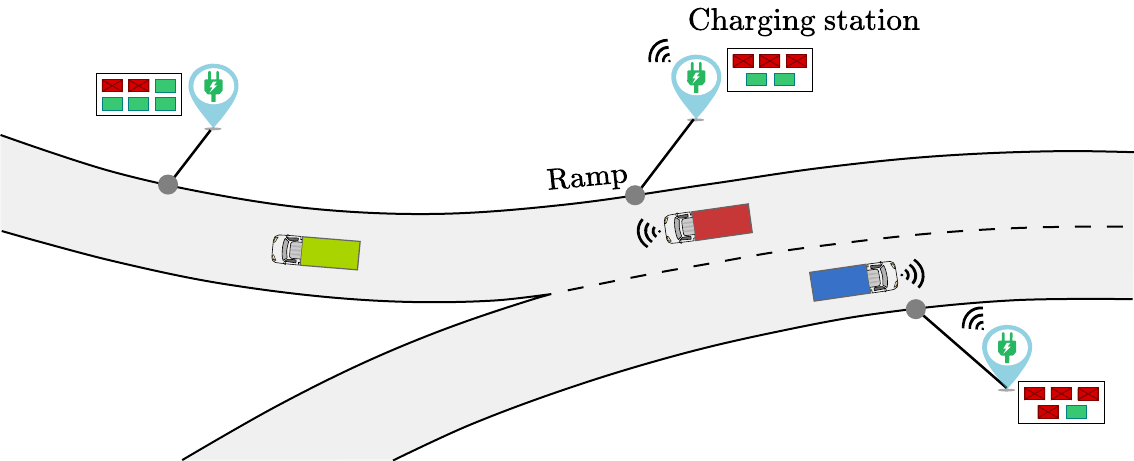}
      \caption{An illustration of the road network and communication scheme between trucks and charging stations. Information changes are triggered between a truck and a charging station when the truck reaches a ramp. Each station has limited charging resources, where the occupied charging ports are shown in red blocks while the available ones are shown in green.}
      \vspace{-15pt}
      \label{Fig.1}
   \end{figure}
   
With the conditions stated in mind, we are now ready to describe our distributed coordination framework. As illustrated in Fig.~\ref{Fig.1}, the main roads of the traffic network are divided into road segments by \emph{ramps}, which are the starting points of shortest detours leading to single charging stations. Whenever a truck reaches a ramp in its pre-planned route, the following sequence of communications and computations occur:  
\begin{itemize}
    \item[1)] The truck sends to the corresponding station the anticipated arrival time, which is the present time plus the detour time $d$ for reaching the station.
    \item[2)] The charging station computes the anticipated waiting time $\Tilde{w}$ and return it to the truck.
    \item[3)] Based on the waiting time $\Tilde{w}$, the truck updates its charging plan and sends to the corresponding station the planned charging time at that station.
    \item[4)] If the planned charging time is nonzero, the charging station places the truck into the queue and updates related information.
\end{itemize}

\begin{figure}[t!]
     \centering
     \includegraphics[width=0.8\linewidth]{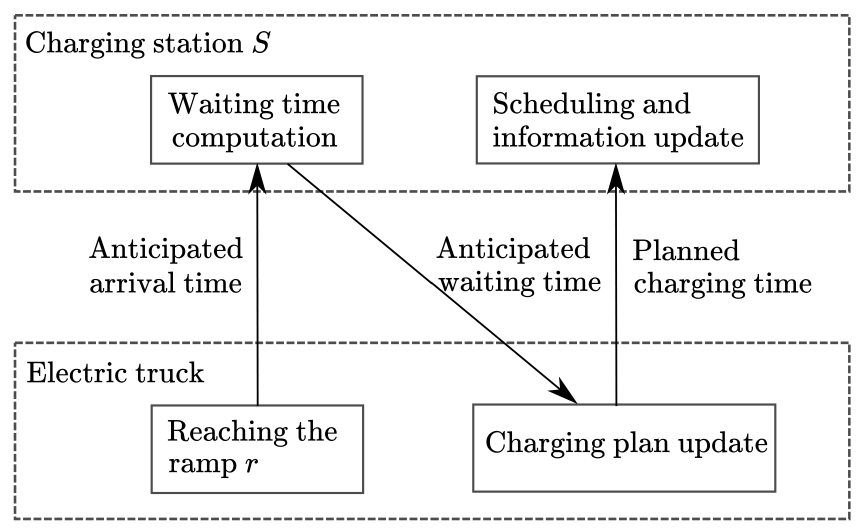}
      \caption{Station-truck coordination framework. Upon arriving at a ramp, the chain of communication and computation is initiated by the truck via sending anticipated arrival times to the corresponding station. The procedure is complete after the station receives the planned charging time from the truck.}
      \vspace{-18pt}
      \label{Fig.2}
   \end{figure}

A summary of the architecture is illustrated in Fig.~\ref{Fig.2}. Note that there is no communication between stations and between individual trucks; there is only information exchange between trucks and stations. Moreover, since a ramp corresponds to a unique station, a truck only communicates with a single station when updating its charging plan. Besides, one may note all the computations are carried out locally, so the coordination framework is fully distributed. In what follows, we provide details on the computations carried out by stations and trucks, respectively.

\section{Scheduling and Computation at Charging Station}\label{Section III}
This section describes the mechanism according to which each charging station schedules charging and updates its local information. For simplicity, we focus on the case where a station only maintains communication with a single truck at any time instance. Extensions to the setup where a station can communicate with a group of vehicles simultaneously are provided in the end. 

Given a charging station $S$, let us denote by $C$ the collection of its charging ports. For every charging port $c\!\in\! C$, it maintains an available time $a_c$, which is the earliest time from when the port becomes available onward. Upon receiving an anticipated arrival time $\Tilde{t}_a$ of a truck, the station computes the anticipated waiting time $\Tilde{w}$ as
\begin{equation}
    \label{eq:waiting_time}
    \Tilde{w}=\max\Big\{\min_{c\in C}(a_c-\Tilde{t}_a),0\Big\},
\end{equation}
which compares the available times of its ports with the received arrival time. The obtained value is sent to the truck that initiated the communication.

According to the designed mechanism, the truck updates its charging plans and sends back to the station the planned charging time $t$. If $t\!=\!0$, the availability times $a_c$ remain unchanged for all $c$. Otherwise, let us denote by $c^*$ the port that achieves the minimum in \eqref{eq:waiting_time}, i.e.,
$$c^*\in \arg\min_{c\in C}a_c.$$
Then the port $c^*$ is assigned to the truck in communication from $\Tilde{t}_a+\Tilde{w}$ for a duration of $t$, and the available time of the port $c^*$ is updated to $a_{c^*}\!=\!\Tilde{t}_a+\Tilde{w}+t$, while other times remain unchanged. 

Note that the mechanism introduced above follows the first-come, first-served rule. It ensures that the charging time is spent consecutively in the same port, and is easy to carry out. On the other hand, it admits extensions provided more information is exchanged between the station and the truck. For example, in addition to the anticipated arrival time $\Tilde{t}_a$, if the truck also provides an anticipated charging time upper bound $\Bar{t}$ (which could be the charging time needed for the battery to be fully charged), the anticipated charging waiting time may be reduced, at the expense of maintaining a detailed schedule of individual ports.\footnote{Note that such an extension is useful only at charging stations reached from several different ramps. If the station is connected by a single ramp, this extension is redundant.} In addition, it may also communicate simultaneously with a group of trucks by providing the waiting times of one port to one truck, according to a certain order. In the present work as well as the subsequent empirical studies, we focus on the case where a station talks to one truck at one time.

\section{Charging Planning by Trucks}\label{Section IV}
Having described the scheduling mechanism of charging stations, we now introduce the computations carried out by individual trucks. As discussed earlier, when reaching a ramp, a truck sends its anticipated arrival time to the corresponding station, initiating communications and computations. This time is simply the sum of the current clock time and the time needed to reach the station from the ramp. In contrast, the computation of the planned charging time at the station requires consideration of various factors. We cast the charging problem addressed by the trucks as mixed-integer linear programs, which are modified in part from our previous work \cite{10147895}. In what follows, we provide a brief account of the model for the charging problems faced by trucks, as well as the optimization problem upon receiving the anticipated waiting time at the present station. For simplicity, constraints related to drowsy driving discussed in \cite{10147895} are not considered here. However, they can be incorporated into the present framework without modification of any part discussed earlier. Interested readers are referred to \cite{10147895} for further details.

\subsection{Model of the Charging Problem}
\begin{figure}[t!]
     \centering
     \includegraphics[width=0.98\linewidth]{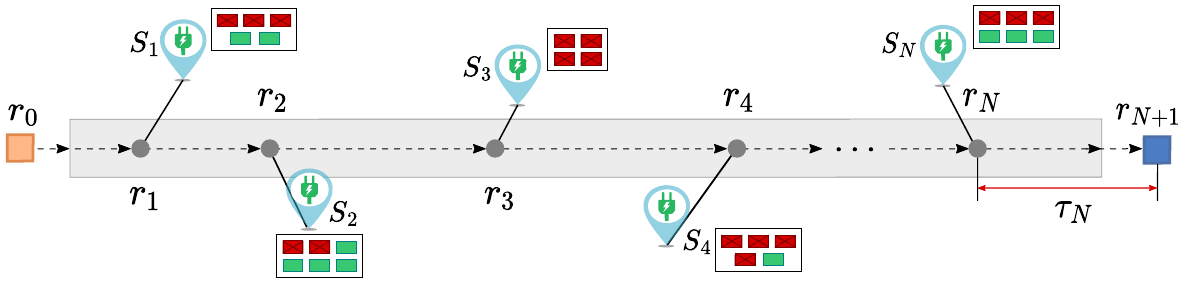}
      \caption{The route model of each truck, where charging stations along the route are denoted by $S_k$, $k\!=\!1,\dots,N$, and ramps leading the shortest detour to each station are denoted by $r_k$ with $k\!=\!1,\dots,N$. The origin and destination are denoted by $r_0$ and $r_{N+1}$, respectively. }
      \vspace{-15pt}
      \label{Fig.3}
   \end{figure}
We assume each truck has a fixed pre-planned route between its origin and destination, as illustrated in Fig.~\ref{Fig.3}. Along this route, $N$ charging stations are available, and there are detours connecting the $k$th station $S_k$ to corresponding ramps $r_k$, $k\!=\!1,\dots,N$. In particular, the origin and destination are denoted by $r_0$ and $r_{N+1}$, respectively. The main route is divided into $N\!+\!1$ road segments by the ramps, with the traverse time from $r_k$ to $r_{k+1}$ denoted by $\tau_k$, $k=0,\dots,N$. The detour time between $r_k$ and $S_k$ is denoted by $d_k$.  

Upon arrival at $r_k$, the following two decisions need to be made: (i) whether to charge the truck at $S_k$; and (ii) how long to charge the truck if deciding to charge at $S_k$. The two decisions are represented by the following variables:
\begin{equation}
    \label{eq:variable}
    b_k\!\in\!\{0,1\},\;t_k\!\in\!\Re_+,\quad k\!=\!1,\dots,N,
\end{equation}
where $b_k\!=\!1$ if the truck decides to charge at its $k$th charging station and $0$ otherwise, $t_k$ denotes the planned charging time of the truck at $S_k$ if $b_k\!=\!1$, and $\Re_+$ represents the set of nonnegative reals.

For the completion of delivery missions, it is necessary to maintain sufficient energy levels upon reaching all ramps. We use hard constraints to encode such requirements. For a given truck, let $e_{\text{ini}}$ be the initial battery at its origin and $e_k$ its remaining battery when it first arrives at its ramp $r_k$. The dynamics of $e_k$ over ramps are given as
\begin{subequations}
\begin{align}
&e_1\!=\!e_{\text{ini}}-\bar{P}\tau_0,\label{eq:e1}\\
&e_{k+1}\!=\!e_k+b_k\Delta{e_k}\!-\!\bar{P}(2b_kd_k\!+\!\tau_k),\quad k\!=\!1,\dots,N,\label{eq:ek}
\end{align}
\end{subequations}
where $\bar{P}$ is the truck's battery consumption per travel time unit and $\Delta{e}_k$ denotes the increased battery charged at $S_k$. To ensure that the remaining battery of a truck when it arrives at its ramp $r_k$ is sufficient for reaching $S_k$, we require that
\begin{subequations}
\begin{align}
&e_k\geq{e_s}+\bar{P}d_k, \quad k\!=\!1,\dots,N,\label{eq:ek_dy}\\
&e_{N+1}\geq{e_s},\label{eq:eN_dy}
\end{align}
\end{subequations} 
where $e_s$ is a constant margin for safe operation. The charging process is approximately modeled by a linear model, given as
\begin{equation}
\label{eq:delta}
    \Delta{e_k}=t_k\min\big\{P_k,P_{\max}\big\},\quad k\!=\!1,\dots,N,
\end{equation}
where $P_k$ denotes the charging power provided by $S_k$ while $P_{\max}$ is the maximum charging power acceptable by the truck's battery. Restricted by the battery capacity of the truck, $\Delta{e_k}$ is confined by  
\begin{equation}
\label{eq:ek_con}
    {0}\leq\Delta{e_k}\leq{e_f}\!-\!\big(e_k\!-\!\bar{P}d_k\big),\quad k\!=\!1,\dots,N,
\end{equation}
where $e_f$ is the full battery of the truck.

Typically, delivery missions have deadlines. However, due to constraints of charging resources, it is not possible to ensure the deadline as a hard constraint. We encode the preference of reaching the destination before the deadline as a soft constraint, as we discuss next.

\subsection{Computing Charging Decisions}
When arriving $r_k$ with battery energy $e_k$, upon receiving the anticipated waiting time $\Tilde{w}_k$ from $S_k$, the truck assumes certain waiting times $\hat{w}_\ell$, $\ell=k+1,\dots,N$, for subsequent stations. To compute the charging decisions $b_k$ and $t_k$, one would like to minimize the cost associated with detouring, charging, and waiting. Moreover, we denote by $T_k$ the difference between the delivery deadline and the present time. This is the remaining travel time for charging planning when arriving at $r_k$. Reaching the destination before the deadline is described as a soft constraint. It involves the anticipated overtime $\Delta{T}_k$ at $r_k$ defined as
\begin{align}
\Delta{T}_k=&~b_k(2d_k+t_k+\Tilde{w}_k)+\sum_{\ell=k+1}^{N}\!\!b_\ell\big(2d_\ell+t_\ell+\hat{w}_\ell\big)+\nonumber\\
&\sum_{\ell=k}^N\tau_\ell-T_k,\label{eq:del_T}
\end{align}
which is the difference between the planned total time to complete the mission and the remaining travel time budget according to the deadline. Thus, $\Delta{T}_k\!>\!0$ indicates the violation of the delivery deadline.  

With the charging model and related quantities defined above, the charging problem upon reaching $r_k$ is cast as
\vspace{-3pt}
\begin{align}
\min_{\{(b_\ell,t_\ell)\}_{\ell=k}^N}\;&\!\kappa\bigg(b_k\big(2d_k\!+\!t_k\!+\!\tilde{w}_k\big)+\!\sum_{\ell=k+1}^N\!\!\!b_\ell\big(2d_\ell\!+\!t_\ell\!+\!\hat{w}_\ell\big)\bigg)+\nonumber\\&\sum_{\ell=k}^N\!\epsilon_\ell{b_\ell}{t_\ell}+\max\big\{\rho\Delta{T}_k,0\big\}\label{eq:cost}\\
\mathrm{s.\,t.} ~\;& \eqref{eq:variable}-\eqref{eq:del_T},\nonumber
\end{align}
where $\kappa$ represents the labor cost per time unit, $\epsilon_\ell$ denotes the electricity cost of $S_\ell$ per charging time unit, and $\rho$ is a tuning parameter for encoding the deadline as a soft constraint. Denote by $\{(b_\ell^*,t_\ell^*)\}_{\ell=k}^N$ the values that attain the minimum of the problem, then the anticipated charging time sending to $S_k$ is the product of $b_k^*$ and $t_k^*$.

In our previous work \cite{10147895}, we have developed a rollout-based method for solving the charging problems approximately. It enables real-time solutions with cheap hardware at little loss of solution quality. The problem may also be solved exactly by a dedicated mixed-integer solver after performing a linearization procedure discussed in \cite[Appendix~F]{bai2023}. 

\section{Simulation Studies}\label{Section V}
This section presents the results of our simulation studies, which demonstrate the effectiveness of the proposed coordination scheme. Swedish road maps and realistic transportation data are employed. We begin by introducing the scenario and parameter settings. The code of the implementation is available at \url{https://github.com/kth-tingbai/Distributed-Charging-for-EVs}.
\begin{figure}[t!]
     \centering
     \includegraphics[width=0.77\linewidth]{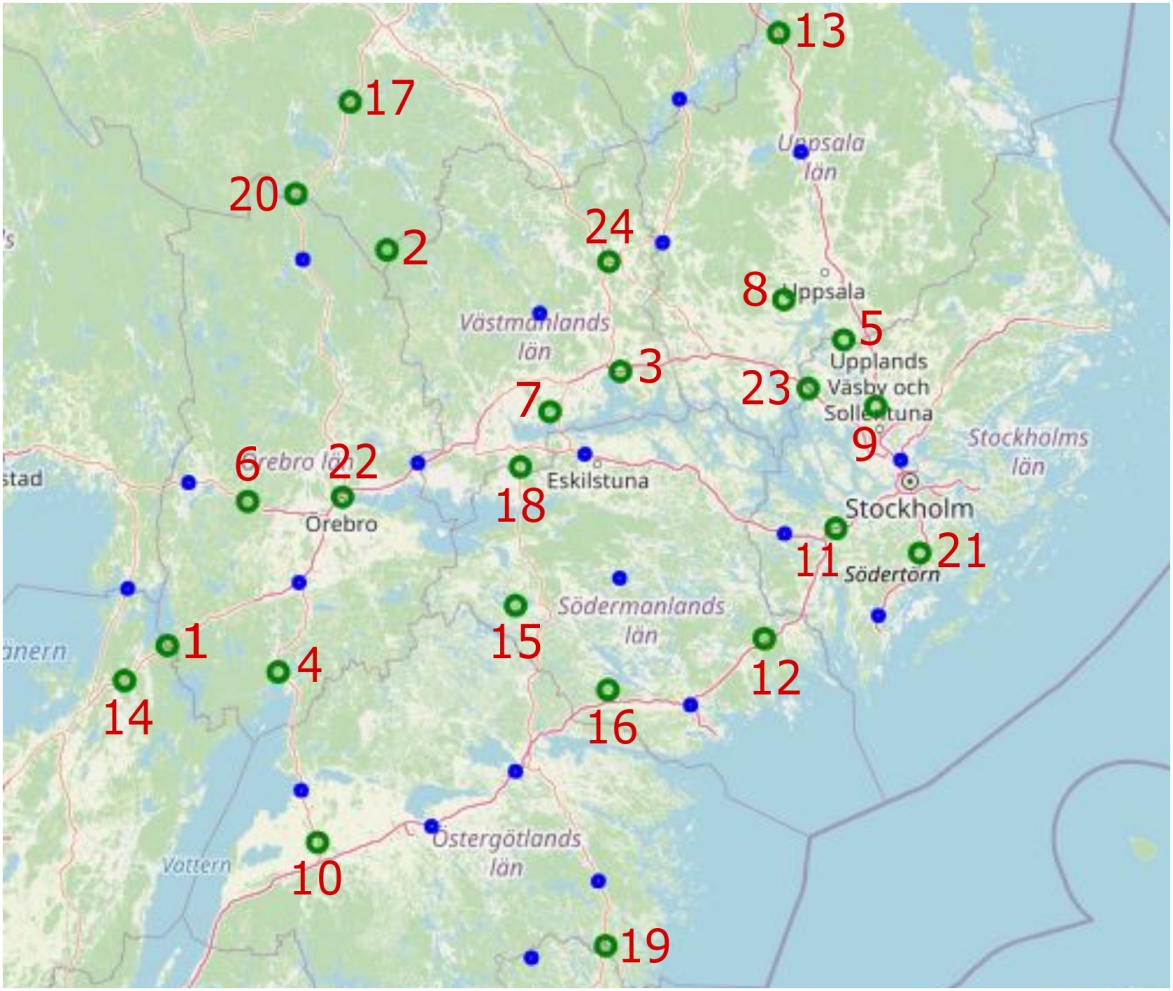}
      \caption{The road network of a region in Sweden, where the potential charging stations are shown by green nodes while hubs, from which the origin and destination pairs are chosen, are shown by blue nodes. Indices of the charging stations are marked in red.}
      \vspace{-15pt}
      \label{Fig.4}
   \end{figure}

\subsection{Scenario and Parameter Settings} We consider a transportation system comprised of $150$ trucks traveling in the Swedish road network, as illustrated in Fig.~\ref{Fig.4}, where nodes shown in blue are hubs from which the origin and destination pair of each truck is selected randomly. Since only a few charging stations that serve commercial electric trucks are in operation today, we use $24$ road terminals attained from the SAMGODS model~\cite{bergquist2016representation} as charging stations in the simulation. The locations of these stations are marked by the green nodes on the map. Trucks' routes are pre-planned using the data from \textit{OpenStreetMap}~\cite{OpenStreetMap}. In addition, with a specified search range, trucks can identify the charging stations along the pre-planned routes. Trucks' travel times on each road segment between ramps and those for detours are obtained accordingly from \textit{OpenStreetMap}.    

We assume trucks start their trips at a random time between 08:00-10:00 a.m., with an arbitrary initial battery over the safe operation bound, i.e., the initial battery is sufficient for the truck to reach the first charging station in its route. The total extra travel time budget of every truck is set as $160$ minutes. This is the total time a truck can use for detours, charging the battery, and waiting in queues before reaching the delivery deadline. Moreover, we assume that every charging station is equipped with $3$ charging ports, each providing charging power of $300$~kWh, and the station follows the first-come, first-served rule to assign charging ports to trucks. Other parameters related to trucks are set in line with the latest published data for electric trucks manufactured by Scania~\cite{ElectricTruck}, as given in Table~\ref{Table1}, where the labor cost is set based on drivers' salaries today in Sweden.

\begin{table}[t]
\caption{Parameter Values} % title name of the table
\vspace{-5pt}
\centering
\begin{tabular}{|c|c|c|c|c|c|} 
\hline
& & & & &\\[-1.5ex]
  \raisebox{1.0ex}{~$P_{\max}$~}&\raisebox{1.0ex}{$\bar{P}$}& \raisebox{1.0ex}{~~~$e_f$~~~}& \raisebox{1.0ex}{~~~$e_s$~~~}&\raisebox{1.0ex}{~~~$\epsilon_{\ell}$~~~} &\raisebox{1.0ex}{~~~$\kappa$~~~}
\\ [-0.5ex]
\raisebox{1.3ex}{kW}&\raisebox{1.3ex}{kWh/min}& \raisebox{1.3ex}{kWh}& \raisebox{1.3ex}{kWh} &\raisebox{1.3ex}{\texteuro/kWh}&\raisebox{1.3ex}{\texteuro/min}
\\[-0.5ex]
\hline % inserts single-line
& & & & &\\[-0.7ex]
 \raisebox{1.2ex}{$375$}&\raisebox{1.2ex}{$1.83$}& \raisebox{1.2ex}{$624$} & \raisebox{1.2ex}{$156$}  & \raisebox{1.2ex}{$0.36$}& \raisebox{1.2ex}{$0.4$}\\[-0.2ex]
\hline% inserts single-line
\end{tabular}
\label{Table1}
\end{table}

\subsection{Strategy Comparison and Evaluation}
\begin{figure}[t!]
     \centering
     \includegraphics[width=0.9515\linewidth]{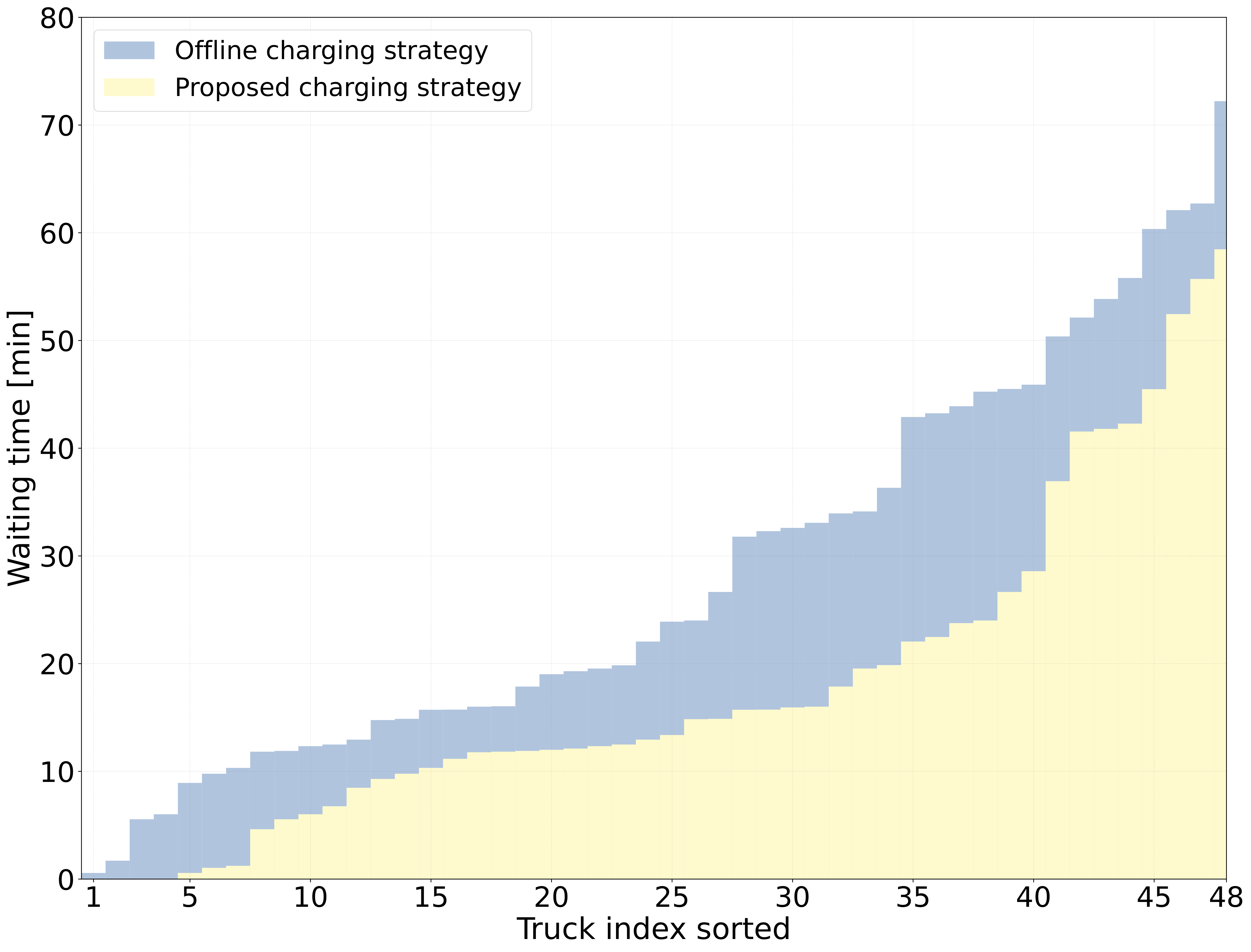}
     \vspace{-5pt}
      \caption{Real waiting times of individual trucks.}
      \label{Fig.5}
   \end{figure}
   
We first compute the optimal charging strategy for individual trucks in an offline manner under the assumption of unlimited charging facilities at stations. Subsequently, we apply the proposed charging scheduling approach to each truck. Upon arriving at a ramp, each truck solves the problem (\ref{eq:cost}) in real-time, leveraging the anticipated waiting time provided by the charging station. Then it updates its charging plan at current and future charging stations and applies that only at the current station. Trucks' real waiting times in their entire trips using the two charging strategies are then evaluated. On average, the computation for an individual truck at one ramp takes $0.1034$ seconds on a laptop with an Intel Core$^{\text{TM}}$ i7-8665U CPU operating at 1.90GHz with 4 cores and 8 logical processors. The waiting times at future stations are decided through trial and error, and $12$ minutes lead to the best overall performance in our tests. 

Fig.~\ref{Fig.5} illustrates the real waiting times of individual trucks following the charging strategies computed offline and at every ramp. For clarity, we sort truck indices according to their waiting times in the two methods while omitting trucks without waiting. As illustrated in Fig.~\ref{Fig.5}, the offline charging strategy causes $22.67$ hours of waiting time in total for $48$ trucks. The proposed approach reduces this waiting time to $14.30$ hours, saving approximately $37\%$ of time for trucks, at the expense of a little communication between trucks and the charging stations they approach.   
\begin{figure}[t!]
     \centering
     \includegraphics[width=1.0\linewidth]{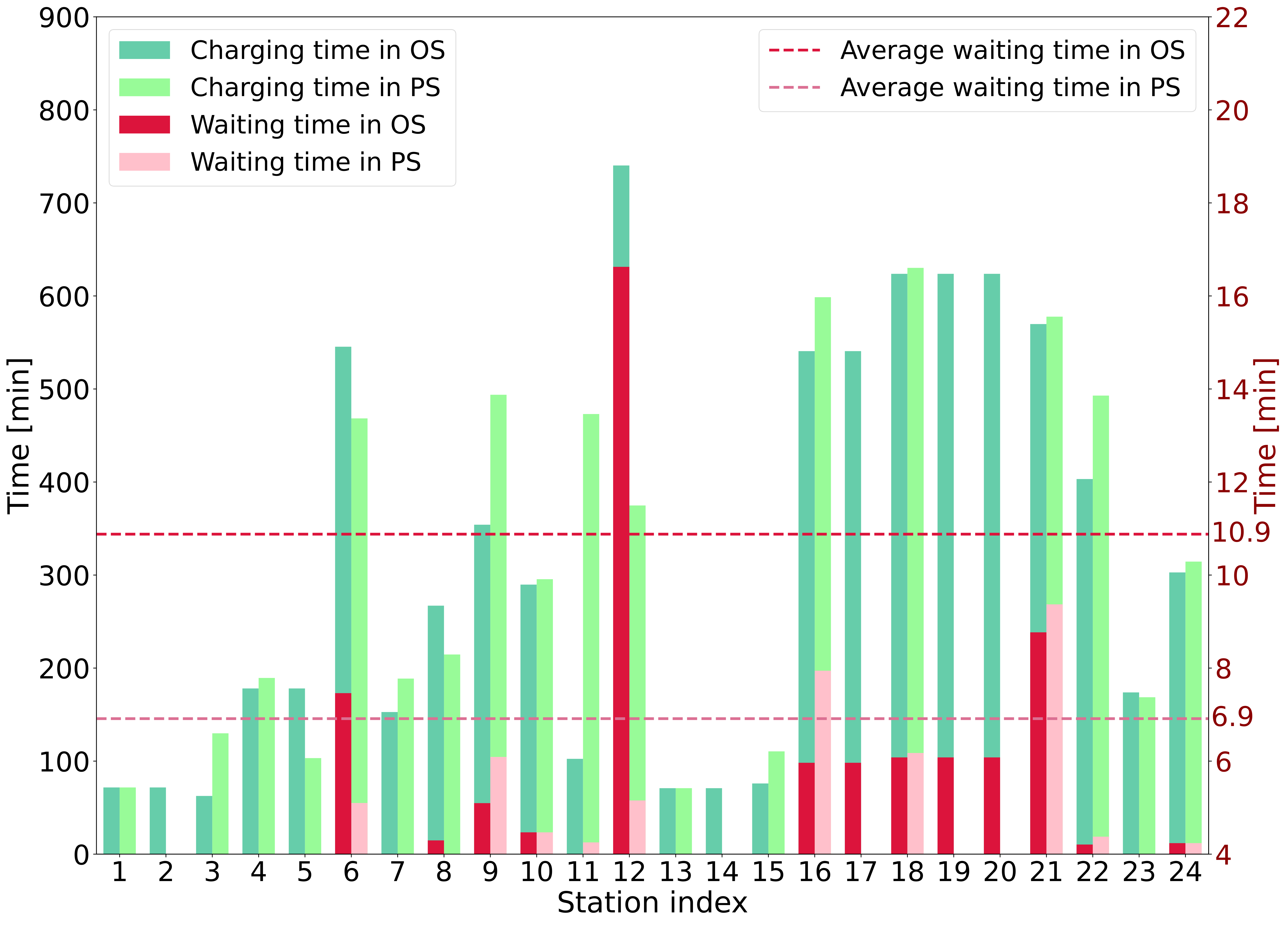}
     \vspace{-16pt}
      \caption{Trucks' charging and waiting times at each station, shown by the bars, and the average waiting times shown by the dashed lines.}
      \label{Fig.6}
   \end{figure}
   
We further evaluate the occupancy of charging ports and show in Fig.~\ref{Fig.6} the total charging and waiting times of all trucks at every charging station. On the horizontal axis, the station indices match those in Fig.~\ref{Fig.4}. For simplicity, the offline strategy and the proposed strategy are denoted by OS and PS, respectively. The simulation results indicate that the proposed coordination scheme alleviates the systemic charging congestion by reducing trucks' average waiting time at a station from $10.9$ to $6.9$ minutes. It is also worth noting that, while our method improves the charging efficiency on a holistic level, it may result in longer waiting times for some trucks due to the diverse routing, battery levels, and travel time schedules of individual trucks.

\begin{figure}[t!]
     \centering
     \includegraphics[width=0.9515\linewidth]{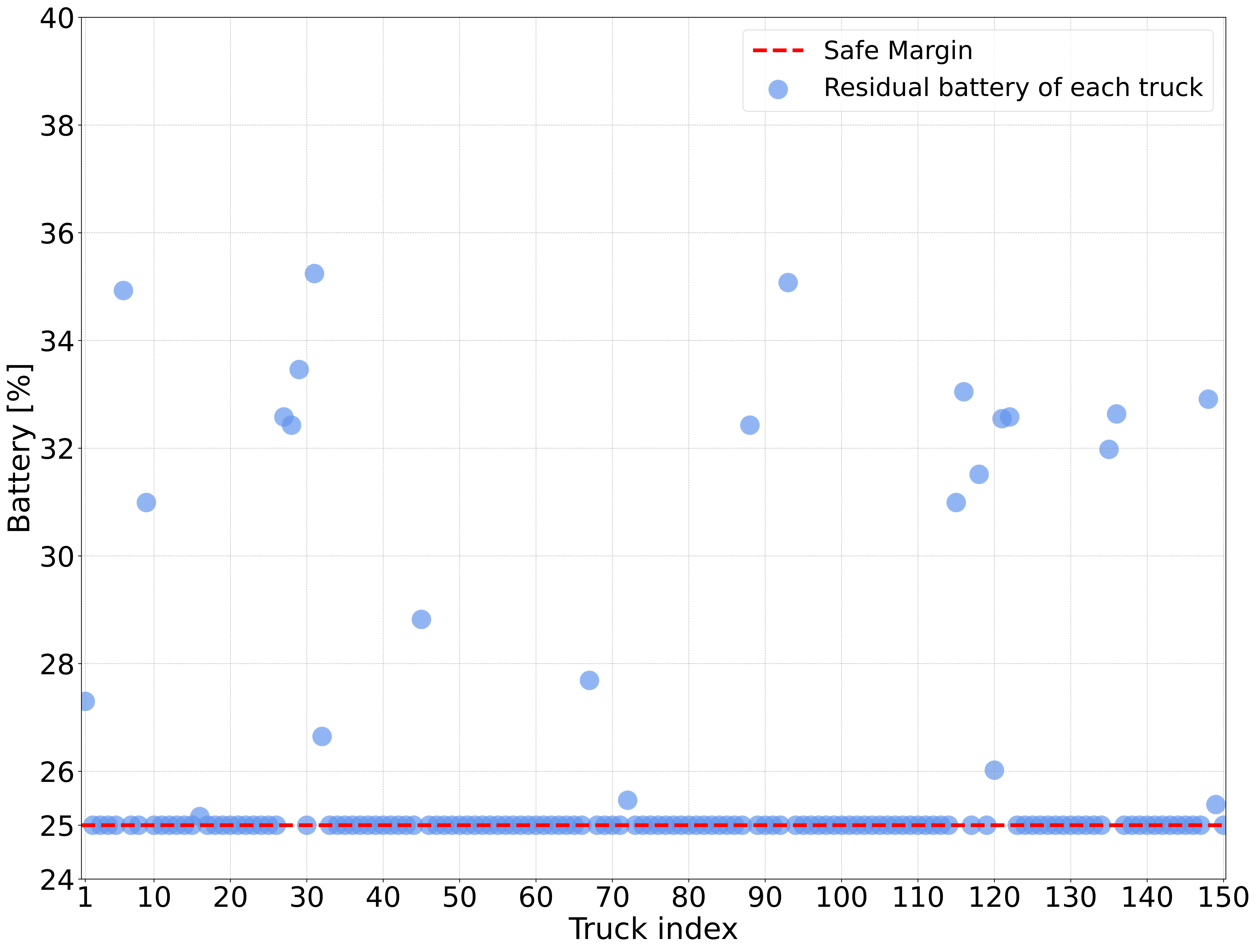}
     \vspace{-5pt}
      \caption{Residual battery upon arriving at the destination.}
      \label{Fig.7}
   \end{figure}

\begin{figure}[!t]
    \centering
    \subfigure[Charging time scheduling at Station 12]{
    \includegraphics[width=8.4cm,height=4.2cm]{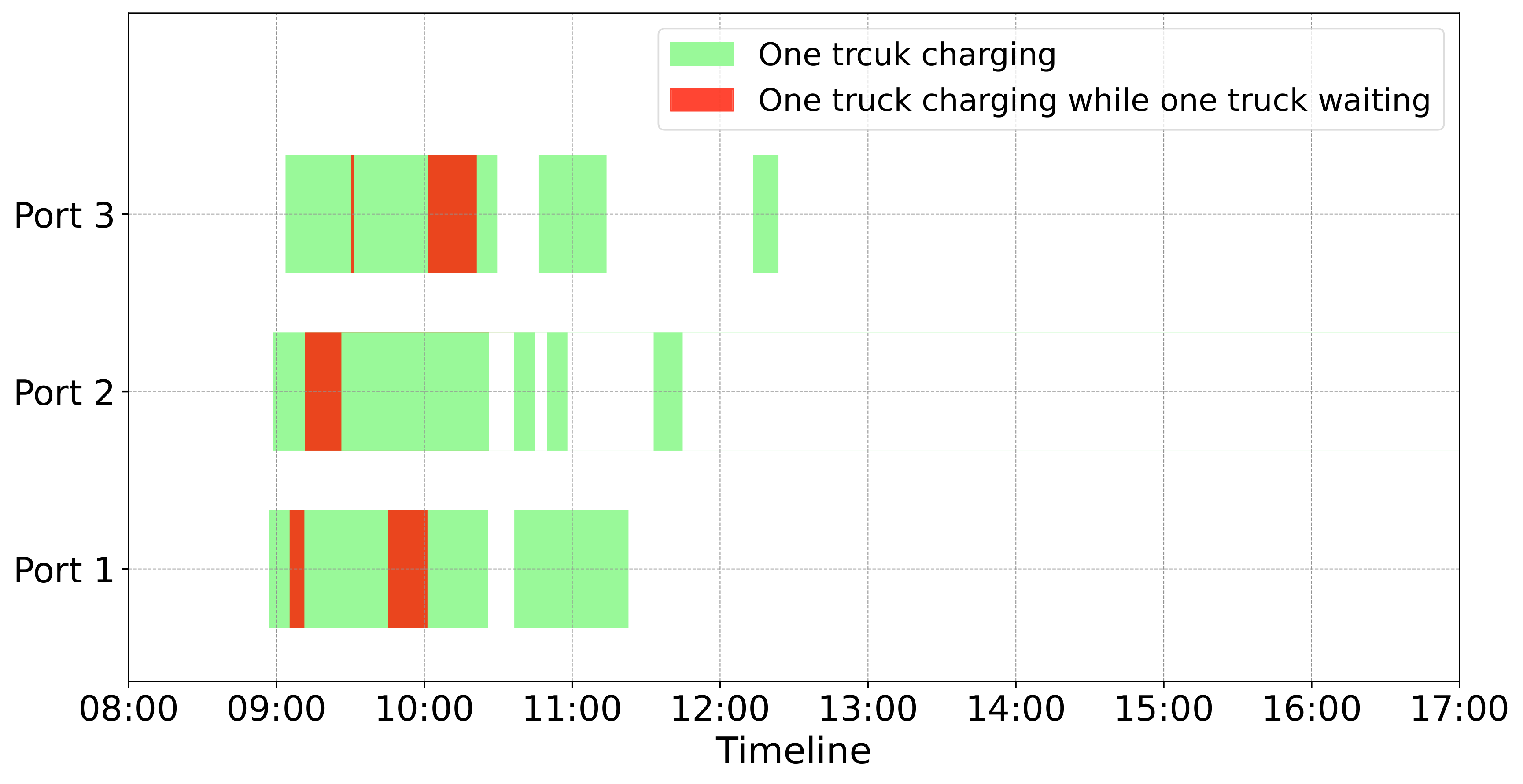}
    }
    \centering
    \subfigure[Charging time scheduling at Station 22]{
    \includegraphics[width=8.4cm,height=4.2cm]{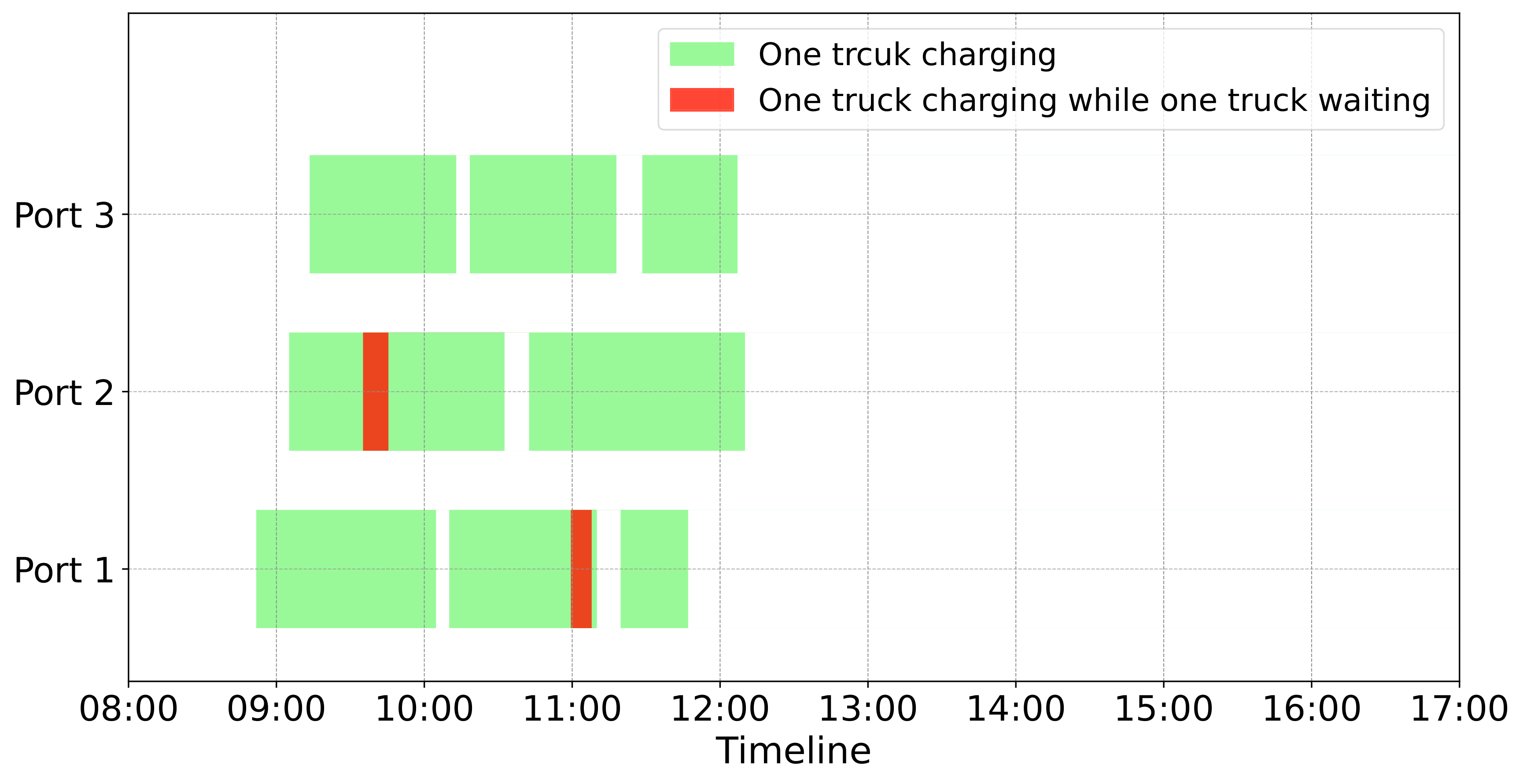}
    }
    \vspace{-6pt}
    \DeclareGraphicsExtensions.
    \caption{Charging ports assignment at Stations 12 (a) and 22 (b).}
    \label{Fig.8}
\end{figure}

In Fig.~\ref{Fig.7} we present the residual battery of every truck upon reaching their destinations, where the proposed charging coordination scheme is applied. As given in Table~\ref{Table1}, the least amount of battery for safe operation is considered as $25\%$ of the full battery (i.e., $(156\!\times\!{100})/624\!=\!25\%$). Fig.~\ref{Fig.7} illustrates that most trucks complete their delivery missions while consuming all the usable batteries. This is because we consider that charging at home is normally cheaper than at a station and the total charging cost during the trip is minimized in addressing (\ref{eq:cost}). It should be noted that our method can easily be modified to handle scenarios where higher residual batteries are required at destinations, for instance, for follow-up deliveries. Finally, Fig.~\ref{Fig.8} shows trucks' charging time scheduling at each charging port at two stations. It follows the first-arrive, first-served assigning rule, and the proposed coordination approach is applied.

\section{Conclusions}\label{Section VI}
In this work, we have studied the charging coordination problem, where a collection of trucks aims to plan where and how long to charge along pre-planned routes for their own benefits, while the charging resources are limited. To facilitate the charging planning of trucks and to mitigate congestion in charging stations, we have proposed a two-layer distributed coordination framework. Built on our earlier work, the trucks compute their charging plans by solving mixed-integer optimization problems, and the stations schedule charging orders according to the first-come, first-served rule. Our scheme only requires simple information exchange between trucks and stations, allows fully distributed computations carried out by them, and enables re-planning by trucks upon approaching each station. We performed numerical studies based on the Swedish road network and compared the results with offline computed charging plans. It was shown that the framework brought great benefits in reducing the collective costs of all parties involved. In future work, we would like to integrate real-time crowd estimation in charging stations to the framework introduced here.     
%%%%%%%%%%%%%%%%%%%%%%%%%%%%%%%%%%%%%%%%%%%%%%%%%%%%%%%%%%%%%%%%%%%%%%%%%%%%%%%%
%\section*{Acknowledgement}
%The authors would like to thank Albin Engholm for providing the simulation data from the SAMGODS model.

%%%%%%%%%%%%%%%%%%%%%%%%%%%%%%%%%%%%%%%%%%%%%%%%%%%%%%%%%%%%%%%%%%%%%%%%%%%%%%%%
\bibliographystyle{IEEEtran}
\bibliography{ECC}

\end{document}